# Elastic Constants and Charge Ordering in α'-NaV$_2$O$_5$

[1]H. Schwenk, [1]S. Zherlitsyn, [1]B. Lüthi, [2]E. Morre, [2]C. Geibel

[1]Physikalisches Institut, Universität Frankfurt, Robert-Mayer Strasse 2-4, D-60054 Frankfurt
[2]Max Planck Institut für chemische Physik fester Körper, Bayreutherstrasse 40, D-01187 Dresden

We present the temperature dependence of shear and longitudinal elastic constants in α'-NaV$_2$O$_5$. For the longitudinal $c_{22}$ and $c_{33}$ modes we find anomalies at $T_c$ in contrast to the Spin Peierls substance CuGeO$_3$ where only the longitudinal mode along the chain shows a pronounced effect at $T_{SP}$. The $c_{66}$ shear mode (propagation along the chain in b-direction polarization in a-direction) shows strong softening of 12%. Such a large effect is absent for all shear modes in CuGeO$_3$. We can interpret this softening with a coupling of the $e_{xy}$ symmetry strain to the charge fluctuation of B$_{1g}$ symmetry. We give the possible low temperature charge distribution

Isotropic one dimensional $S=1/2$ Spin chains are gapless and show no magnetic ordering. But if there are two different magnetic positions for the ions in the magnetic unit cell they may show a singlet triplet gap. Examples are spin ladders, alternating spin chains and Spin Peierls substances.

The susceptibility of NaV$_2$O$_5$ can be explained in the high temperature phase by a Bonner-Fisher law with antiferromagnetic interaction typical for one dimensional spin systems [1]. The broad maximum is around 350K, corresponding to a coupling constant $J$=220K. At 34K a structural phase transition occurs and the susceptibility drops. This was interpreted as a Spin-Peierls phase transition [1,2]. Neutron scattering [2,3], magnetic susceptibility [1,4], magnetic resonance ESR [5,6,7,8] and specific heat [9] gave evidence of a singlet - triplet splitting of the spin ½ chains and a lattice deformation below $T_c$. The crystal structure was believed to be the non centro symmetric space group $P2_1mn$ with two different V-ion positions. This implies that there exists already a charge ordered phase at room temperature. In this case the V$^{4+}$ ($S$=1/2) ions form a spin chain along the b-direction separated from each other by nonmagnetic chains consisting of V$^{5+}$($S$=0) ions. However NMR measurements have shown that there is only one vanadium position at room temperature while in the low temperature phase there are two different V positions [10]. Recently several redeterminations of the crystal structure [11, 14, 22] gave a high temperature centro-symmetric space group $Pmmn$ ($D^{13}_{2h}$) with all crystallographic vanadium positions being identical. This means that the average valence of the V-ions is 4.5 in the high temperature phase. A charge ordering into V$^{4+}$ and V$^{5+}$ below $T_c$ should occur as deduced also from NMR measurements [10]. This transition can be accompanied or followed by a structural transition as perhaps observed by thermal expansion experiments [12]. There are various proposals for the charge ordering [13, 14, 15, 16] none of them giving a simple linear chain of V$^{4+}$ below $T_c$. Most of them favour a zig-zag spin chain. There is also a crystallographical analyses that predicts a low dimensional phase with V$^{4+}$ double chains isolated by V$^{5+}$ chains [16].

In this investigation we present sound velocity results for α'-NaV$_2$O$_5$ above and below the structural transition. Contrary to previous experiments [17] we investigate not only longitudinal modes but also transverse ones. This enables us to reach conclusions about the charge ordering below $T_c$. We find that the transverse $c_{66}$ ($q\cong0$) mode exhibits strong softening due to a direct coupling to charge fluctuations. We find a bilinear coupling of the corresponding strain mode $e_{xy}$ to the charge order parameter which describes the temperature dependence of the $c_{66}$ mode perfectly. A bilinear coupling of a $q\cong0$ mode to a Spin Peierls order parameter is not possible because the Spin Peierls transition is a $q=\pi/a$ effect.

We used LiNbO$_3$ transducers for exciting elastic waves in the frequency range 10 - 450 MHz. Most measurements were performed at 150 MHz. The relative sound velocity and attenuation was measured using a pulsed transmission technique by a phase comparison method [18]. The sample was placed in a Styropor box. No sound is propagating through this material. Two samples were measured. The sizes were about 2x5x1mm$^3$ for the $a,b$ and $c$ direction. The transducers were fixed on both sites of the sample with organic liquid 'Thiokol 32'. By measuring also the absolute value of the sound velocity $v$ one can calculate the elastic constant $c_{ij}=\rho v^2$ where $\rho$ is the mass density of the crystal. With this formula it is also possible to calculate the relative change of the elastic constant $\Delta c/c_0=2\Delta v/v_0+(\Delta v/v_0)^2$. The quadratic term is only important for large changes of the elastic constant as e.g. for the $c_{66}$ mode.

In fig.1 we show relative sound velocity data as a function of temperature for longitudinal waves in the $b$- ($c_{22}$ mode) and $c$- ($c_{33}$ mode) direction. The elastic constant $c_{ii}$ is gained from the corresponding sound velocity $v$. The anomaly at $T_c$ is of the order of 0.1-0.2 %. The temperature dependence of the $c_{22}$ mode agrees with previous measurements [17]. In the inset of Fig.1 we show the detailed temperature dependence of the $c_{33}$ mode in the vicinity of the transition $T_c$. Apart from the pronounced anomaly at $T_c \cong 33.2$K we find additional kinks in the curve for $T > T_c$ which could have the same origin as mentioned before for the thermal expansion experiments [12]. Note that the crystal is from the same batch for the two experiments. The $c_{22}$ mode shows qualitatively the same behavior as the $c_{33}$ mode except for the structures shown in the inset. We measured the $c_{22}$ mode in a crystal from a different batch. Therefore the existence of multiple phase transitions has to be investigated in much more detail. In the inorganic Spin-Peierls CuGeO$_3$ only the longitudinal mode along the chain shows such a dip like anomaly as observed in NaV$_2$O$_5$ but not the mode perpendicular to the chain. This difference between the two substances will be emphasized even more when considering the shear modes.

The really exciting results are given in Fig.2 where we plot the relative sound velocities of the $c_{66}$ and the $c_{55}$ shear modes as a function of temperature. We notice a very strong softening of the $c_{66}$ mode of 7 % from 75K to $T_c$ almost two order of magnitudes larger than the longitudinal modes discussed above. This mode belongs to the strain $e_{xy}$ and has propagation along the chain axis $b$ and polarization along the a axis or vice versa. Clearly it has the characteristics of a soft mode. The other mode $c_{55}$ and similar modes for CuGeO$_3$ (fig. 2b) exhibit only small anomalies of the order of 0.1%.

We interpret our results using the Landau theory of phase transition. We define the charge density $\rho(T)=\rho_0+\Delta\rho(T)$. This must be invariant under all symmetry operations in the high temperature phase. Therefore $\Delta\rho$ is zero for the high temperature phase and starts growing below. $\Delta\rho$ can be written as $\Delta\rho = \sum_i \eta_i j_i$ where $\eta_i$ are the order parameters and $j_i$ are the irreducible representations of $D_{2h}$. We assume that we have only one order parameter and call it $\eta$. The free energy density can than be expanded in powers of $\eta$. In addition the free energy has also a strain - order parameter coupling term $F_c$: $F=F_0+\alpha\eta^2+\beta\eta^4+F_c+F_{el}$. Here $F_0$ is the part of the free energy that stays constant through the phase transition, $F_{el}=c_{ij}e_{ij}^2/2$ is the elastic energy and $F_c$ can be bilinear in the symmetry strain $e_\Gamma$ and the order parameter $\eta_\Gamma$: $F_c = ge_\Gamma\eta_\Gamma$. Here $g$ is the phonon order parameter coupling constant. This coupling will lead to a softening of the corresponding sound wave mode with [19,20]

$$c(T)=c_0(T-T_o)/(T-\Theta) \quad (1)$$

where $\Theta$ is the phase transition without the strain - order parameter coupling and $T_o = \Theta + g^2/2\alpha'c_0$ with $\alpha = \alpha'(T - \Theta)$ the coefficient of the first term in the Landau expansion. $T_o$ indicates the phase transition temperature if the mode is completely soft [20]. A bilinear coupling means that both the strain and the order parameter have the same symmetry and also the same $q$-vector. In our case of ultrasonic waves is $q \cong 0$.

The strain-order parameter coupling can also be quadratic in the order parameter (e.g. for longitudinal modes): $F_c=ge\eta^2$. This will lead to a step like behavior of the elastic constant [20]:

$$c(T)=c_0 - g^2/\beta . \quad (2)$$

In Fig. 3 we show the $c_{66}$ mode as a function of temperature together with a fit of eq. (1) (dashed line). It is seen that the fit is perfect with $\Theta = 30.34$K and $T_o = 30.62$K. The strain order parameter coupling expressed by $T_o - \Theta = 0.28$K is very small. This is evident from the experimental results where the softening starts only for $T < 75$K. This mode has $B_{1g}$ symmetry. This means that there has to be a $q=0$ order parameter with $B_{1g}$ symmetry. For $T < T_c$ domain wall stress effects influence the sound velocity and give strong attenuation effects.

On the other hand as shown in Fig. 2 the $c_{55}$ mode hardly shows any softening, only some structure of < 0.1% in the vicinity of $T_c$ due to interference with the $c_{66}$ mode. The $c_{55}$ mode has $B_{2g}$ symmetry in the high temperature phase.

In Fig. 2b we show the corresponding shear modes for the Spin Peierls compound CuGeO$_3$. Here $c_{55}$ is the mode with propagation along the chains ($c$-axis) and polarization along $a$. It is the analogue to the $c_{66}$ mode in NaV$_2$O$_5$. Unlike this latter mode it shows only an effect of $10^{-4}$ at $T_{SP} = 12K$. This is further strong experimental evidence that in α'-NaV$_2$O$_5$ the $c_{66}$ mode couples bilinearly to the charge fluctuations.

Now we want to give a simple symmetry argument to substantiate the bilinear coupling for the $c_{66}$ mode and the quadratic coupling for the longitudinal modes. A complete symmetry analysis has to include the space group $D_{2h}^{13}$ which is beyond the scope of the paper. In Fig. 4 we show the arrangement of the V-ions in the $a$-$b$ plane together with the unit cell. If we take the charges $r_1$ - $r_4$ of the V-ions as indicated in the figure (the other charges can be reached with a simple translation) the symmetry elements of $D_2$ acting on these $r_i$ form a reducible representation with $G_{red} = G_{A1g} + G_{B1g} + G_{B2g} + G_{B3g}$. We consider only g-representations because of coupling to the strains (which have g-representations). With $B_{1g} \cdot B_{1g} = A_{1g}$ we have therefore a bilinear coupling of the strain mode $e_{B1g} = e_{xy}$ with a charge fluctuation of $B_{1g}$ symmetry as used above. The other possible couplings (with strains $e_{xz}$ and $e_{yz}$) must have smaller coupling constants $g$.

Using the projection operator technique we get basis of the irreducible representation. The basis of the $A_{1g}$ mode is $r(A_{1g}) = r_1 + r_2 + r_3 + r_4$. This is a charge fluctuation mode with equal weight for all atoms which means that all ions have the same charge in the high symmetry phase. The basis of the $B_{1g}$ mode is instead $r(B_{1g}) = (r_1 - r_2 + r_3 - r_4)$. We take the average valence of the V-Ions to 4.5 and $\Delta r = 0.5$. Then we get a charge ordered phase which gives a zig-zag chain as proposed for the low temperature charge ordering by some theories and experiments [13-15]. These are indicated in Fig. 4b.

Assuming that the order parameter has $B_{1g}$ symmetry we find that for all longitudinal elastic modes we can couple quadratically to the order parameter with $(B_{1g} \cdot B_{1g})e_{ii}$ as we have found for the $c_{33}(e_{zz})$ and $c_{22}(e_{yy})$ modes (Fig. 1). The absence of step like anomalies is due to order parameter fluctuations.

Coupling of elastic modes to charge fluctuations is not confined to α'-NaV$_2$O$_5$. Recently the case of Yb$_4$As$_3$ was discussed [19], another case is the $c_{44}$-softening in Magnetite [21].

This research was supported in part by SFB 252. We thank T.Goto for discussions.

**Figures:**

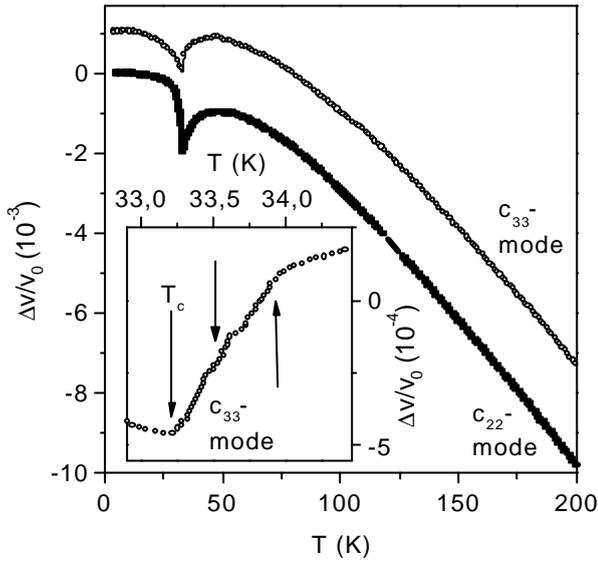

Fig.1: Temperature dependence of the sound velocity of the $c_{22}$ $c_{33}$ $\alpha'$- NaV O$_5$ relative sound velocity in the vicinity of the phase transition. The absolute sound velocity of the $_{22}$ mode is 6500m/s ( = 4.2K).

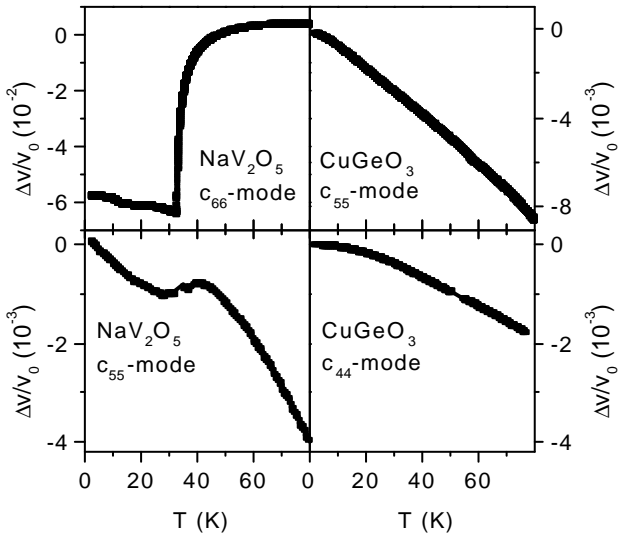

Fig.2 Temperature dependence of the relative shear sound velocities for $\alpha'$-NaV$_2$O$_5$ (Fig. 2a) and CuGeO$_3$ (Fig. 2b). For $\alpha'$- NaV$_2$O$_5$ we show the $c_{55}$ mode and the $c_{66}$ mode (Propagation along the b-axis with polarization along the a-axis). The absolute sound velocities are $v(c_{55}) = 2600$m/s and $v(c_{66}) = 4200$ m/s. For CuGeO$_3$ we show the $c_{44}$ mode and the $c_{55}$ mode (the equivalent mode to the $c_{66}$ mode in NaV$_2$O$_5$).

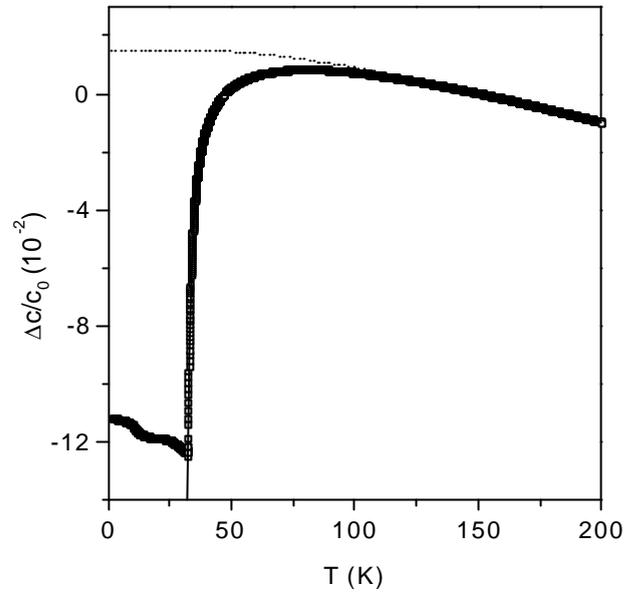

Fig.3 Temperature dependence of the $c_{66}$ mode in $\alpha'$- NaV$_2$O$_5$ together with a fit using Eq.(1) (dashed line). The doted line indicates the background [23].

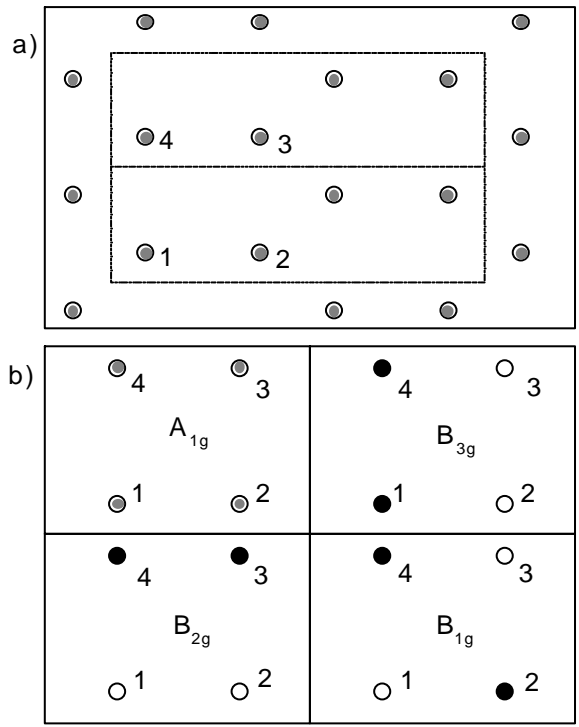

Fig.4: a) *a-b* projection of the V-ions for NaV$_2$O$_5$. Two adjacent unit cells are framed by dashed lines. The 4 V ions considered for the symmetry analysis are numbered.
b) Charge distributions for the different symmetries. Open circles V$^{4+}$, gray circles V$^{4.5+}$, black circles V$^{5+}$. $B_{1g}$ is the symmetry deduced from our sound wave analysis.